\documentclass[11pt,a4paper]{article}
\pdfoutput=1
\usepackage{jcappub}

\newcommand{\Lya}{Ly$\alpha$~}

\title{Broadband distortion modeling in Lyman-$\alpha$ forest BAO fitting}

\author[a]{Michael~Blomqvist,}
\author[a]{David~Kirkby,}
\author[b]{Julian~E.~Bautista,}
\author[c]{Andreu~Arinyo-i-Prats,}
\author[d,e,f]{Nicol\'as~G.~Busca,}
\author[g,c]{Jordi~{Miralda-Escud\'e,}}
\author[h]{An\v{z}e~Slosar,}
\author[i]{Andreu~Font-Ribera,}
\author[a]{Daniel~Margala,}
\author[j,k]{Donald~P.~Schneider}
\author[h]{and Jose~A.~Vazquez}

\affiliation[a]{Department of Physics and Astronomy, University of California, Irvine, 92697, USA}
\affiliation[b]{Department of Physics and Astronomy, University of Utah, Salt Lake City, UT 84112, USA}
\affiliation[c]{
Institut de Ci\`encies del Cosmos,
Universitat de Barcelona, IEEC-UB,
08028 Barcelona, Catalonia
}
\affiliation[d]{
APC, University of Paris Diderot, CNRS/IN2P3, CEA/IRFU, Observatoire de Paris, Sorbonne Paris Cit\'e, F-75205 Paris, France
}
\affiliation[e]{
Observat\'orio Nacional, 
Rua Gal.~Jos\'e Cristino 77, 
Rio de Janeiro, RJ - 20921-400, Brazil
}
\affiliation[f]{
Laborat\'orio Interinstitucional de e-Astronomia, - LIneA, 
Rua Gal.Jos\'e Cristino 77, 
Rio de Janeiro, RJ - 20921-400, Brazil  
}
\affiliation[g]{
Instituci\'o Catalana de Recerca i Estudis Avan\c{c}ats,
08010 Barcelona, Catalonia
}
\affiliation[h]{
Brookhaven National Laboratory, 
Bldg 510, 
Upton, NY 11973, USA 
}
\affiliation[i]{
Lawrence Berkeley National Laboratory, One Cyclotron Road,
Berkeley, CA 94720, USA
}
\affiliation[j]{
Department of Astronomy and Astrophysics, The Pennsylvania State University,
University Park, PA 16802
}
\affiliation[k]{
Institute for Gravitation and the Cosmos, The Pennsylvania State University,
University Park, PA 16802
}

\emailAdd{cblomqvi@uci.edu, dkirkby@uci.edu, bautista@astro.utah.edu, andreuaprats@icc.ub.edu, ngbusca@apc.in2p3.fr, miralda@icc.ub.edu, anze@bnl.gov, afont@lbl.gov, dmargala@uci.edu, dps7@psu.edu, jvazquez@bnl.gov}

\abstract{In recent years, the Lyman-$\alpha$ absorption observed in the spectra of high-redshift quasars has been used as a tracer of large-scale structure by means of the three-dimensional Lyman-$\alpha$ forest auto-correlation function at redshift $z\simeq 2.3$, but the need to fit the quasar continuum in every absorption spectrum introduces a broadband distortion that is difficult to correct and causes a systematic error for measuring any broadband properties. We describe a $k$-space model for this broadband distortion based on a multiplicative correction to the power spectrum of the transmitted flux fraction that suppresses power on scales corresponding to the typical length of a Lyman-$\alpha$ forest spectrum. Implementing the distortion model in fits for the baryon acoustic oscillation (BAO) peak position in the Lyman-$\alpha$ forest auto-correlation, we find that the fitting method recovers the input values of the linear bias parameter $b_{F}$ and the redshift-space distortion parameter $\beta_{F}$ for mock data sets with a systematic error of less than 0.5\%. Applied to the auto-correlation measured for BOSS Data Release 11, our method\newpage \noindent improves on the previous treatment of broadband distortions in BAO fitting by providing a better fit to the data using fewer parameters and reducing the statistical errors on $\beta_{F}$ and the combination $b_{F}(1+\beta_{F})$ by more than a factor of seven. The measured values at redshift $z=2.3$ are $\beta_{F}=1.39^{+0.11\ +0.24\ +0.38}_{-0.10\ -0.19\ -0.28}$ and $b_{F}(1+\beta_{F})=-0.374^{+0.007\ +0.013\ +0.020}_{-0.007\ -0.014\ -0.022}$ (1$\sigma$, 2$\sigma$ and 3$\sigma$ statistical errors). Our fitting software and the input files needed to reproduce our main results are publicly available.}


\begin{document}
\maketitle


\section{Introduction}

Baryon acoustic oscillations (BAO) is one of the most well-established observational probes of cosmic acceleration \cite{2013PhR...530...87W}. The characteristic length scale of BAO imprinted in the large-scale clustering of the density field provides a standard ruler that can be used to study the expansion rate and geometry of the Universe as a function of redshift \cite{2003ApJ...598..720S}. A measurement of the BAO scale in the radial direction probes the Hubble parameter, whereas a measurement in the transverse direction provides constraints on the angular diameter distance. Until recently, the BAO technique had mainly been exploited using the clustering of galaxies at redshift $z<1$ \cite{2005ApJ...633..560E,2005MNRAS.362..505C,2010MNRAS.401.2148P,2011MNRAS.418.1707B,2011MNRAS.416.3017B}. The Baryon Acoustic Oscillation Survey \cite{2013AJ....145...10D} (BOSS) of the Sloan Digital Sky Survey III \cite{2011AJ....142...72E,2006AJ....131.2332G} (SDSS-III) reported a 1\% measurement of the cosmic distance scale at $z=0.57$ and a 2\% measurement at $z=0.32$ \cite{2014MNRAS.441...24A} using a sample of nearly one million galaxies taken from SDSS-III Data Release 11 \cite{2015ApJS..219...12A} (DR11).

The expansive BOSS data set includes an unprecedented number of high-redshift quasar spectra that enable us to explore the large-scale structure at $z>2$ through the Lyman-$\alpha$ (Ly$\alpha$) forest technique. The spatially correlated absorption of \Lya photons by the neutral hydrogen component of the intergalactic medium along the lines of sight to quasars traces the matter distribution, thus allowing for measurements of BAO during the epoch of decelerated expansion when the universe was strongly matter dominated \cite{2003ApJ...585...34M,2007PhRvD..76f3009M}. The first observation of three-dimensional large-scale correlations in the fluctuations of the \Lya forest absorption using BOSS quasar spectra was reported in \cite{2011JCAP...09..001S}, providing observational constraints on the \Lya forest linear bias parameter and redshift-space distortion parameter at $z=2.25$. Using a sample of approximately 50,000 BOSS quasar spectra, the first detection of the BAO peak feature in the \Lya forest auto-correlation function at $z\simeq2.3$ was presented in \cite{2013A&A...552A..96B,2013JCAP...04..026S,2013JCAP...03..024K}. The analysis was improved to yield a more precise measurement of the BAO peak position using the three-times larger DR11 quasar sample \cite{2015A&A...574A..59D}. Complementary measurements of BAO at a similar redshift can be achieved by considering the correlated clustering of the intergalactic medium with the quasars themselves. The three-dimensional large-scale cross-correlation of quasars and the \Lya absorption was detected in \cite{2013JCAP...05..018F} and subsequently extended to larger separations for DR11 to measure the BAO scale \cite{2014JCAP...05..027F}. By combining the BAO results for the DR11 auto- and cross-correlation, the Hubble parameter and the angular diameter distance at $z=2.34$ were measured to 2\% and 3\% accuracy, respectively. The cosmological implications of the BAO measurements from the BOSS galaxy clustering and \Lya forest analyses, in combination with data from other probes, were studied in \cite{2014arXiv1411.1074A}.

A method for fitting \Lya forest BAO was developed in \cite{2013JCAP...03..024K}. Here, we refine the method by constructing a model for the distortions of the broadband shape of the measured correlation function introduced by the analysis. Broadband distortions primarily arise due to continuum fitting in the \Lya forest of quasar spectra. The standard approach for accommodating these distortions in BAO fitting is to introduce a smooth broadband function in $r$-space that is additive and/or multiplicative on the correlation function. The broadband function is typically required to feature many free parameters (usually about ten) in order to provide a good fit to the measured correlation function. In the method developed here, the broadband distortion is instead modeled in $k$-space as a multiplicative correction to the power spectrum of the transmitted flux fraction that suppresses power on scales corresponding to the typical length of the \Lya forest. This distortion model has a clearer physical interpretation and allows for a much more precise determination of the \Lya forest linear bias parameter and redshift-space distortion parameter. This approach additionally provides a better fit to the \Lya auto-correlation using fewer parameters.

Our analysis uses the DR11 \Lya auto-correlation from \cite{2015A&A...574A..59D} based on the BOSS spectra of 137,562 quasars in the redshift range $2.1\leq z\leq3.5$. The BAO fit results reported in that paper can be directly compared to the results obtained here using the distortion model, because changes are only made in the BAO modeling and not in the data. The fit input files and the fitting code are made publicly available, so that readers may reproduce our main results.

This paper is organized as follows: Section~\ref{sec:baomodel} describes the BAO model, including models of the non-linear correction of the transmission power spectrum. In section~\ref{sec:distmodel}, we define our model of the broadband distortion due to continuum fitting. Section~\ref{sec:baofitting} explains the method for fitting the BAO model to the measured auto-correlation function. In section~\ref{sec:mocks}, we test the distortion model using auto-correlation functions for mock data sets, and section~\ref{sec:results} describes the fits to the DR11 auto-correlation and presents our main results. The paper concludes with a discussion in section~\ref{sec:discussion}.

\section{BAO model}
\label{sec:baomodel}

In this section, we describe our method to calculate a theoretical prediction of the \Lya auto-correlation that can be fitted to the measured auto-correlation to determine the BAO scale. The modeling is done in $k$-space by means of the \Lya forest transmission power spectrum, which is Fourier transformed into the \Lya auto-correlation. Our fiducial cosmology is the same as in \cite{2015A&A...574A..59D}, namely a spatially flat $\Lambda$CDM universe with $\Omega_{\rm m}=0.27$, $h=0.7$, $\Omega_{\rm b}h^{2}=0.0227$, $\Omega_{\rm \nu}h^{2}=0.00064$ (massive neutrinos), $\sigma_{8}=0.79$ and $n_{s}=0.97$.

\subsection{Transmission power spectrum}
\label{subsec:fluxpower}

We model the power spectrum of the \Lya forest transmitted flux fraction at redshift $z$, including linear redshift-space distortions, non-linear effects and broadband distortion due to continuum fitting, as
\begin{equation}
P_{F}(k,\mu_{k},z) = b_{F}^{2}(z)\left[1 + \beta_{F}\mu_{k}^{2}\right]^{2}P_{\rm NL}(k,\mu_{k},z)D_{\rm NL}(k,\mu_{k})D_{\rm C}(k_{\parallel}) \ ,
\label{eq:Pk}
\end{equation}
where $k$ is the modulus of the wavenumber, $\mu_k$ is the cosine of the angle of the wavenumber from the line of sight, and $k_{\parallel}=k\mu_{k}$ is the component along the line of sight. The model builds on the linear theory prediction (e.g., \cite{2003ApJ...585...34M}), for which $P_{F}$ is reduced to the redshift-dependent bias factor $b_{F}(z)$ of the \Lya transmission, the term depending on the redshift-space distortion factor $\beta_{F}$ (which in principle depends also on redshift, but is assumed to be constant in this paper), and the linear matter power spectrum. Here, we separate the non-linear effects modifying the power spectrum into a multiplicative non-linear correction for small scales, $D_{\rm NL}(k,\mu_{k})$, that is described in section~\ref{subsec:nlcorr}, and a smoothing term that models the effect of large-scale bulk motions on the BAO peak described below. The term $D_{\rm C}(k_{\parallel})$ models the broadband distortion due to continuum fitting, and is described in detail in section~\ref{sec:distmodel}.

The non-linear matter power spectrum is modeled by
\begin{equation}
P_{\rm NL}(k,\mu_{k},z) = \exp\left[-k^{2}\Sigma^{2}(\mu_{k})/2\right]P_{\rm L}(k,z)\ ,
\label{eq:PkNL}
\end{equation}
where $P_{\rm L}(k,z)$ is the (isotropic) linear matter power spectrum obtained from CAMB \cite{2000ApJ...538..473L} for our fiducial cosmology. Non-linear effects on the matter density field due to large-scale bulk velocity flows are modeled with an anisotropic Gaussian roll-off of the linear matter power spectrum \cite{2007ApJ...664..675E}, with
\begin{equation}
\Sigma^2(\mu_{k})=\mu_{k}^2\Sigma_{\parallel}^2+(1-\mu_{k}^2)\Sigma_{\perp}^2\ .
\end{equation}
The effect of the anisotropic non-linear broadening is a convolution of the correlation function that broadens the BAO peak. The parameters quantifying the strength of the radial and transverse broadening are related by
\begin{equation}
\frac{\Sigma_{\parallel}}{\Sigma_{\perp}}=1+f\ ,
\end{equation}
where $f=d(\ln g)/d(\ln a)\approx\Omega_{\rm m}^{0.55}(z)$ \cite{2007APh....28..481L} and $g(z)$ is the growth factor. We adopt the values $\Sigma_{\parallel}=6.41$~Mpc/$h$ and $\Sigma_{\perp}=3.26$~Mpc/$h$ expected at $z=2.4$ \cite{2013JCAP...03..024K}. The non-linear broadening model provides a valid description of the effect on scales of the BAO feature, but applies unphysical filtering of small-scale structure. A more refined model would feature scale dependent broadening parameters.

The redshift evolution of the linear matter power spectrum at $z>2$ (where $f\approx1$) is governed by the growth factor with respect to a reference redshift $z_{\rm ref}$ such that
\begin{equation}
P_{\rm L}(k,z) = P_{\rm L}(k,z_{\rm ref})g^{2}(z)=P_{\rm L}(k,z_{\rm ref})\left( \frac{1+z_{\rm ref}}{1+z} \right)^{2}\ .
\label{eq:PLevo}
\end{equation}
Following the result of \cite{2006ApJS..163...80M}, we assume that $P_{F}$ evolves with redshift according to the combination
\begin{equation}
b_{F}^{2}(z)g^{2}(z) = b_{F}^{2}(z_{\rm ref})\left( \frac{1+z}{1+z_{\rm ref}} \right)^{3.8}\ .
\label{eq:PFevo}
\end{equation}
Combining equation~(\ref{eq:PLevo}) and equation~(\ref{eq:PFevo}) produces the result that the linear bias evolves as
\begin{equation}
b_{F}(z) = b_{F}(z_{\rm ref})\left( \frac{1+z}{1+z_{\rm ref}} \right)^{2.9}\ .
\end{equation}
We will use the shorthand notation $b_{F}\equiv b_{F}(z_{\rm ref})$ in the remainder of this paper.

The three-dimensional \Lya forest auto-correlation function $\xi_{F}(\vec{r},z)$ is most generally related to $P_{F}(\vec{k},z)$ by a Fourier transform
\begin{equation}
\xi_{F}(\vec{r},z) = \frac{1}{(2\pi)^{3}} \int d^{3}\vec{k} P_{F}(\vec{k},z) \exp(-i \vec{k}\cdot\vec{r})\ ,
\end{equation}
which can be evaluated numerically with a three-dimensional Fast Fourier Transform (3D FFT). Specializing to power spectra that are cylindrically symmetric, and therefore only functions of $k_{\parallel}$ and $k_{\perp}$, the polar angle integration can be performed explicitly,
\begin{equation}
\xi_{F}(r_{\parallel},r_{\perp},z) = \frac{1}{2\pi} \int_0^{\infty} d k_{\perp} k_{\perp} J_{0}(k_{\perp} r_{\perp}) F(r_{\parallel},k_{\perp},z)\ ,
\end{equation}
where $F(r_{\parallel},k_\perp,z)$ is the 1D Fourier transform of $P_{F}$ along $k_{\parallel}$ only
\begin{equation}
F(r_{\parallel},k_{\perp},z) \equiv \frac{1}{2\pi} \int_{-\infty}^{+\infty} d k_{\parallel} P_{F}(k_{\parallel},k_{\perp},z) \exp(-i k_{\parallel} r_{\parallel})\ .
\end{equation}
This result suggests an alternative two-step approach to the numerical computation: first, perform $N_{\parallel}N_{\perp}$ 1D FFTs, followed by $N_{\parallel}N_{\perp}$ separate 1D numerical integrations, where $N_{\parallel}$ and $N_{\perp}$ are the number of grid points along the radial and transverse axis, respectively. We are currently using the 3D FFT method, for which we extract $\xi_{F}(r_{\parallel},r_{\perp},z)$, and leave a two-step implementation for future work. Our 3D FFT uses grid spacing $\Delta r=4$~Mpc/$h$ and $N=400$ grid points along each axis, corresponding to coverage up to $k_{\rm max}=\pi/\Delta r=0.79$~$h$/Mpc with spacing $\Delta k=2\pi/(N\Delta r)=3.9\times10^{-3}$~$h$/Mpc. This resolution is sufficient to reach a satisfactory accuracy of the correlation function calculation when fitting for the BAO peak position.

\subsection{Non-linear correction}
\label{subsec:nlcorr}

The function $D_{\rm NL}$ in equation~(\ref{eq:Pk}) represents a correction due to non-linear effects that are dependent on the non-linear clustering of matter, and the hydrodynamic and thermal evolution of intergalactic gas. These non-linear effects can only be properly modeled through hydrodynamic simulations of the intergalactic medium. A parameterized formula for the non-linear correction $D_{\rm NL}(k,\mu_{k})$ can then be constructed by fitting to the ratio of the calculated $P_{F}$ for the simulation to $P_{\rm L}$. We note here that the anisotropic non-linear broadening that we include in the Gaussian term in equation~(\ref{eq:PkNL}) arises from fluctuations at the BAO scale, which are not captured in the simulations that have been used so far for calculating $D_{\rm NL}$ because of their small box size. We can therefore safely assume that the non-linear broadening is a separate effect from anything that is already included in $D_{\rm NL}$, although this will not be true with future simulations with larger box size. In this paper, we employ two non-linear correction models from the literature for separate parts of the analysis.

The first model, which we designate DNL1, is the fitting formula of  \cite{2003ApJ...585...34M},
\begin{equation}
D_{\rm NL}(k,\mu_{k})=\exp\left[\left(\frac{k}{k_{\rm NL}}\right)^{p_{\rm NL}}-\left(\frac{k}{k_{\rm P}}\right)^{p_{\rm P}}-\left(\frac{k\mu_{k}}{k_{\rm V}(k)}\right)^{p_{\rm V}}\right]\ ,
\end{equation}
where
\begin{equation}
k_{\rm V}(k)=k_{\rm V0}\left( 1+\frac{k}{k_{\rm V}^{\prime}} \right)^{p_{\rm V}^{\prime}}\ .
\end{equation}
The DNL1 model has eight parameters and we use the fixed values from table~1 of \cite{2003ApJ...585...34M} derived for $b_{F}^{2}=0.0173$ and $\beta_{F}=1.58$ at $z=2.25$: $k_{\rm NL}=6.40$~$h$/Mpc, $p_{\rm NL}=0.569$, $k_{\rm P}=15.3$~$h$/Mpc, $p_{\rm P}=2.01$, $k_{\rm V0}=1.220$~$h$/Mpc, $p_{\rm V}=1.50$, $k_{\rm V}^{\prime}=0.923$~$h$/Mpc and $p_{\rm V}^{\prime}=0.451$. The DNL1 model is used exclusively in our tests on mock data sets, as explained in section~\ref{sec:mocks}. Since these parameter values are consistently used for both generating and analyzing the mock data sets, it makes no difference that they were derived for a different fiducial cosmology.

The second model, denoted DNL2, is the fitting formula of \cite{2015arXiv150604519A},
\begin{equation}
D_{\rm NL}(k,\mu_{k})=\exp\left[q_{\rm NL}\Delta^{2}(k)\left(1-\left(\frac{k}{k_{\rm V}}\right)^{a_{\rm V}}\mu_{k}^{b_{\rm V}}\right)-\left(\frac{k}{k_{\rm P}}\right)^{2}\right]\ ,
\end{equation}
where
\begin{equation}
\Delta^{2}(k)=\frac{k^{3}P_{\rm L}(k)}{2\pi^{2}}\ .
\end{equation}
The DNL2 model has five parameters and depends on the linear matter power spectrum $P_{\rm L}(k)$. We use the DNL2 model in our fits to the DR11 auto-correlation in section~\ref{sec:results}. Table~\ref{table:dnl2} lists the parameters $q_{\rm NL}$, $k_{\rm V}$, $a_{\rm V}$, $b_{\rm V}$ and $k_{\rm P}$ as a function of redshift; we interpolate linearly in each parameter to the effective redshift of the BAO measurement. The simulations of \cite{2015arXiv150604519A} feature values of the cosmological parameters that are consistent with Planck results \cite{2014A&A...571A..16P}, which makes transferring the derived DNL2 parameters to our fiducial cosmology non-optimal. However, the differences are small once the amplitude of the linear matter power spectrum is re-scaled by the factor $(0.834/0.79)^2$ to match their value $\sigma_{8}=0.834$.

\begin{table}
\begin{center}
\begin{tabular}{|l|ccccc|}
\hline
$z$ & $q_{\rm NL}$ & $k_{\rm V}$ & $a_{\rm V}$ & $b_{\rm V}$ & $k_{\rm P}$ \\
\hline
3.0 & 0.792 & 1.16 & 0.578 & 1.63 & 17.1 \\
2.8 & 0.773 & 1.16 & 0.608 & 1.65 & 19.1 \\
2.6 & 0.781 & 1.15 & 0.611 & 1.64 & 21.0 \\
2.4 & 0.851 & 1.06 & 0.548 & 1.61 & 19.5 \\
2.2 & 0.867 & 1.05 & 0.514 & 1.60 & 19.3 \\
\hline
\end{tabular}
\end{center}
\caption{Parameter values as a function of redshift for the non-linear correction model DNL2. The parameters $k_{\rm V}$ and $k_{\rm P}$ are given in units of $h$/Mpc.}
\label{table:dnl2}
\end{table}

Both model parameterizations feature three terms inside the exponential function which represent different physical effects: the isotropic enhancement of power due to non-linear growth (subscript ``NL"); the isotropic suppression of power due to gas pressure below the Jeans scale (subscript ``P"); and the suppression of power due to line-of-sight non-linear peculiar velocity and thermal broadening (subscript ``V").

\begin{figure}
   \begin{center}
   \includegraphics[width=3.5in]{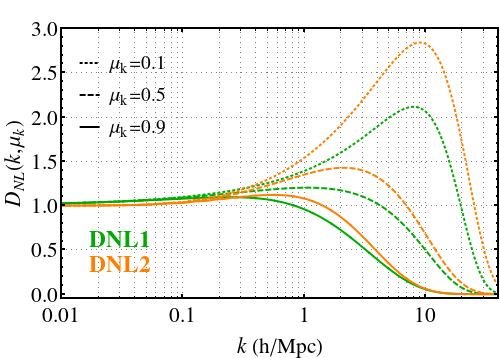}
   \caption{Illustration of the non-linear correction models DNL1 (green) and DNL2 (orange). The solid, dashed and dotted lines show the models for $\mu_{k}=0.9$, $\mu_{k}=0.5$ and $\mu_{k}=0.1$, respectively. DNL1 adopts the parameter values at $z=2.25$ specified in the text. For DNL2, we have used parameter values at $z=2.3$: $q_{\rm NL}=0.859$, $k_{\rm V}=1.055$~$h$/Mpc, $a_{\rm V}=0.531$, $b_{\rm V}=1.605$, $k_{\rm P}=19.4$~$h$/Mpc.}
   \label{fig:nlcorrection}
   \end{center}
\end{figure}

Figure~\ref{fig:nlcorrection} illustrates the non-linear correction models implemented for analyzing the mock data sets (DNL1) and the DR11 data (DNL2). DNL1 is fixed at $z=2.25$, while DNL2 is shown for $z=2.3$, closer to the effective redshift of the DR11 BAO measurement. 
The fact that there are some differences between the models is not surprising, because they were derived using somewhat different techniques for simulations with different properties and fiducial cosmologies. The most important qualitative difference between the models is that DNL2 quickly converges to unity for decreasing $k$, as one would expect in the linear regime, whereas DNL1 converges more slowly and thus produces a non-linear contribution even on large scales. In our application, the 3D FFT resolution used for the BAO fits restricts the wavenumber range to $k<0.79$~$h$/Mpc.

\begin{figure}[htb]
   \begin{center}
   \includegraphics[width=6.0in]{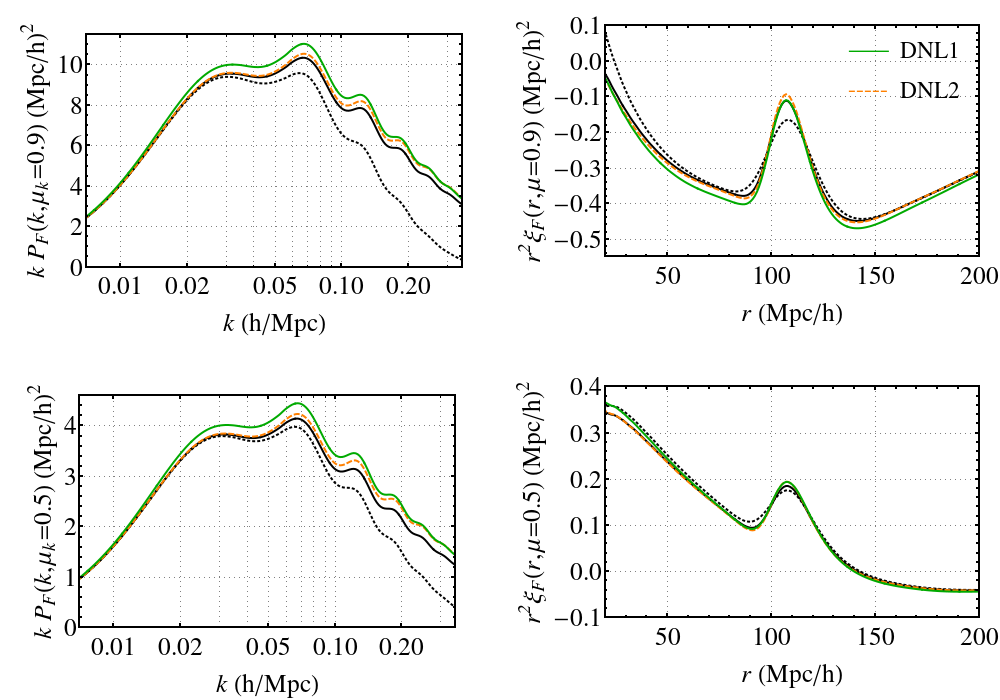}
   \caption{Effects on the $k$-weighted power spectrum $P_{F}$ (left-hand panels) and corresponding $r^2$-weighted correlation function $\xi_{F}$ (right-hand panels) from the non-linear contributions at $z=2.3$. Top panels show the results for $\mu=0.9$ and bottom panels for $\mu=0.5$. Curves indicate the results for no non-linear effects (solid, black), defined by $\Sigma_{\parallel}=\Sigma_{\perp}=0$ and $D_{\rm NL}=1$; only including the anisotropic non-linear broadening (dotted, black); only including the non-linear correction model DNL1 (solid, green); only including the non-linear correction model DNL2 (dashed, orange). We have adopted $\beta_{F}=1.4$ and $b_{F}(1+\beta_{F})=-0.351$ (corresponding to $b_{F}(1+\beta_{F})=-0.336$ at $z=2.25$ \cite{2011JCAP...09..001S}). Distortion due to continuum fitting has not been included ($D_{\rm C}=1$).}
   \label{fig:nlpowercorrelation}
   \end{center}
\end{figure}

Figure~\ref{fig:nlpowercorrelation} shows the effects of the non-linear correction and the anisotropic non-linear broadening on $P_{F}$ and the corresponding correlation function at $z=2.3$. The increase of power on large and intermediate scales by DNL1 affects the broadband shape of the correlation function. In comparison, DNL2 produces a modest change of the shape. The suppression of power on small scales by the non-linear broadening smooths the correlation function to modify the broadband shape and broaden the peak.

\section{Distortion model}
\label{sec:distmodel}

Fitting the continuum level in the \Lya forest of quasar spectra is an operation that induces distortions of the broadband shape of the correlation function. Our analysis is based on the DR11 auto-correlation for which the \Lya forest continua were fitted using the ``C2" method \cite{2015A&A...574A..59D}. In this method, the \Lya forest continuum $C_{q}$ for each quasar is modeled as the product of the mean continuum $\bar{C}$ for the quasar sample, determined by stacking the normalized spectra in the rest frame, and a wavelength-dependent function that sets the overall continuum amplitude and slope,
\begin{equation}
C_{q}(\lambda)=\left[a_{q}+b_{q}\log(\lambda)\right]\bar{C}(\lambda_{\rm rf})\ .
\end{equation}
The parameters $a_{q}$ and $b_{q}$ are fitted using a maximum likelihood approach that imposes that the transmitted flux fraction $F$, calculated as the ratio of the observed flux with the continuum, follows an assumed log-normal probability distribution function \cite{2013A&A...552A..96B}. The spatially correlated fluctuations in the transmission field are calculated as
\begin{equation}
\delta_{F}(\lambda)=\frac{F(\lambda)}{\bar{F}(\lambda)}-1\ , 
\end{equation}
where the mean transmitted flux fraction $\bar{F}$ is determined by calculating the average of all $F$ values at each wavelength, such that, by construction, the mean value of $\delta_{F}$ vanishes at each wavelength. Broadband distortions arise because the continuum fitting cannot distinguish between a change in the continuum amplitude and a true offset in $\delta_{F}$, and will typically favor the former to adjust out the large-scale Fourier modes of $\delta_{F}$. The exact characteristics of the broadband distortion depend on the details of the continuum fitting method as well as the properties of the quasar sample, meaning that the distortion will be different if one changes the fitting method or modifies the sample.

Analytically modeling the broadband distortion introduced by continuum fitting is complicated, because a large number of quasars with different properties and over a wide redshift range are used to measure the correlation function. There can be intrinsic variations in the quasar spectral shape and flux calibration systematic errors, and the distortion depends on how these variations are allowed in the method to fit the continuum. In addition, the physical length of the \Lya forest changes with redshift, with weights at each spectral pixel that vary depending on each particular quasar observation. The approach that has been taken so far is to use analytical
prescriptions for estimating an additive distortion correction to the correlation function \cite{2011JCAP...09..001S,2012JCAP...11..059F}, but this has suffered from modeling uncertainties that degrade the accuracy to measure large-scale, broadband properties of the correlation function.

In this paper, we adopt a different approach based on noticing two facts about the distortion correction. First, in $k$-space, the correction should be a multiplicative function that depends only on the radial component $k_{\parallel}$. Second, this function should suppress the large-scale power on the scale that corresponds to the typical length of the \Lya forest in our quasar spectra. To derive the first property of the distortion correction, we consider the impact of this distortion on single Fourier modes $\delta_{F}(k_{\parallel},k_{\perp})$ of a fixed $k_{\parallel}$ but different $k_{\perp}$, approximating the quasar spectra to be along nearly parallel lines of sight. The variations in $k_{\perp}$ are only affecting each individual spectrum by a phase shift, but this phase shift is uniformly distributed from 0 to $2\pi$ independently of the value of $k_{\perp}$. Since the continuum is determined for each individual spectrum without any reference to nearby parallel spectra, the average suppression of the power can only depend on $k_{\parallel}$. The correction is multiplicative since a change in the undistorted model (e.g., a change in $b_{F}$) should scale correspondingly in the distorted model. The second property is a consequence of fitting the continuum on each spectrum over the length in which the \Lya forest absorption is observed, which sets the characteristic scale determining which Fourier modes are removed most effectively.

\begin{figure}[htb]
   \begin{center}
   \includegraphics[width=3in]{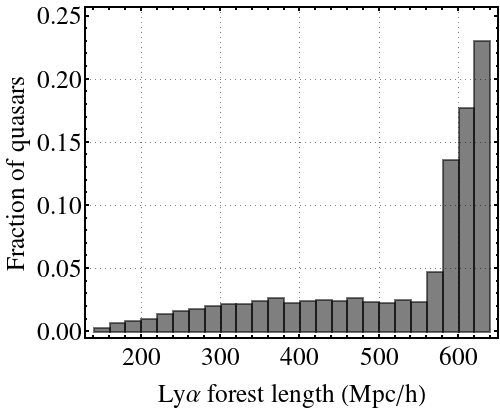}
   \caption{Distribution of the \Lya forest lengths for the BOSS DR11 quasars in the redshift range $2.1\leq z\leq3.5$ that were used to measure the auto-correlation. The adopted bin size is 20~Mpc/$h$. The calculated forest lengths are the maximum comoving separations given the quasar redshift and the forest rest-frame wavelength range $1040<\lambda_{\rm rf}<1200$~\AA. The wavelength cut at observed wavelength $\lambda=3600$~\AA, near the edge of the spectrograph, truncates the spectra of quasars with $z<2.48$.}
   \label{fig:forestlength}
   \end{center}
\end{figure}

The DR11 auto-correlation is based on the spectra of 137,562 quasars in the redshift range $2.1\leq z\leq3.5$. Using the rest-frame wavelength range $1040<\lambda_{\rm rf}<1200$~\AA~as the width of the \Lya forest, the maximum forest length (i.e. the comoving separation spanned by the first and last spectral pixel of the forest) is approximately 635 Mpc/$h$ at $z=2.47$. The length decreases at lower redshift, because the spectra are truncated at observed wavelength $\lambda=3600$~\AA~near the edge of the BOSS spectrograph \cite{2013AJ....146...32S}. At higher redshift, the length decreases slowly because of the increasing Hubble parameter. Figure~\ref{fig:forestlength} displays the distribution of the forest lengths for the quasar sample, with a mean value of approximately 520 Mpc/$h$. The scale of the typical forest length corresponds in $k$-space to $k_{\parallel}\sim0.01$~$h$/Mpc, so we expect our distortion function to vary on this characteristic scale.
 
We construct the distortion model $D_{\rm C}(k_{\parallel})$ as a sigmoid-type function that is zero at $k_{\parallel}=0$ and smoothly increases to approach unity on small scales. We consider two parameterizations and compare their performances using fits to auto-correlation functions for mock data sets in section~\ref{subsec:mockfits}. The first model, denoted DC1, is given by
\begin{equation}
D_{\rm C}(k_{\parallel})=\tanh\left[\left(\frac{k_{\parallel}}{k_{\rm C}}\right)^{p_{\rm C}}\right]\ ,
\end{equation}
and the second model, designated DC2, is
\begin{equation}
D_{\rm C}(k_{\parallel})=\left[\frac{\left(k_{\parallel}/k_{\rm C}+1\right)^{3/2}-1}{\left(k_{\parallel}/k_{\rm C}+1\right)^{3/2}+1}\right]^{p_{\rm C}}\ .
\end{equation}
Both distortion models have two free parameters, $k_{\rm C}$ and $p_{\rm C}$, that are fitted simultaneously with the parameters of the BAO model. Because of the differences in parameterizations, the meanings of $k_{\rm C}$ and $p_{\rm C}$ are not exactly the same between the models. As described in section~\ref{subsec:mockfits}, we fix the $p_{\rm C}$ parameter for DC2 based on the fit results for mock data sets, in order to neutralize its degeneracy with $k_{\rm C}$.

\begin{figure}[htb]
   \begin{center}
   \includegraphics[width=3.0in]{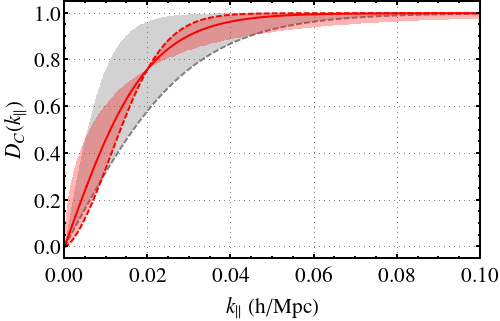}
   \includegraphics[width=3.0in]{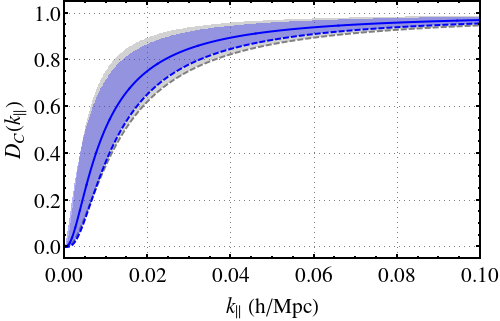}
   \caption{Illustration of the distortion models. Left: Distortion model DC1 for parameter values  $k_{\rm C}=0.02$~$h$/Mpc and $p_{\rm C}=1$ (solid, red). The gray and red bands are the regions enclosed by changing the $k_{\rm C}$ and $p_{\rm C}$ parameters by $\pm$50\% (the gray and red dashed lines indicate the larger values). Right: Distortion model DC2 for parameter values $k_{\rm C}=0.003$~$h$/Mpc and $p_{\rm C}=3$ (solid, blue). The gray and blue bands are the regions enclosed by changing the $k_{\rm C}$ and $p_{\rm C}$ parameters by $\pm$50\% (the gray and blue dashed lines indicate the larger values).}
   \label{fig:distmodel}
   \end{center}
\end{figure}

\begin{figure}[htb]
  \begin{center}
   \includegraphics[width=6.0in]{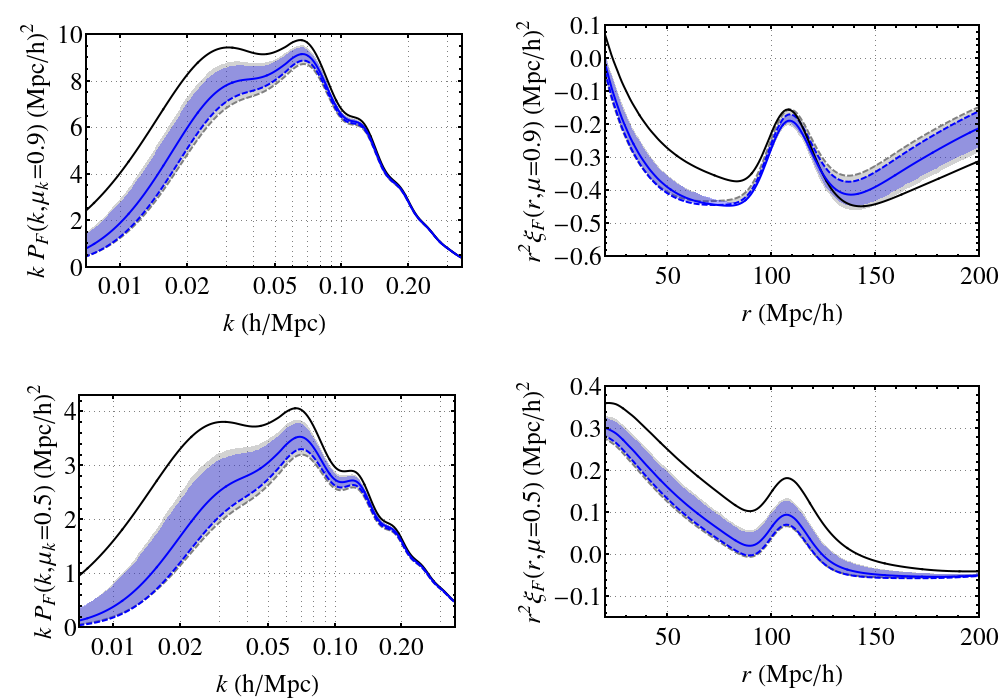}
   \caption{Result of applying the distortion model DC2 to the $k$-weighted power spectrum $P_{F}$ (left-hand panels) and corresponding $r^2$-weighted correlation function $\xi_{F}$ (right-hand panels) at $z=2.3$. Top panels show the results for $\mu=0.9$ and bottom panels for $\mu=0.5$. Curves indicate the undistorted (solid, black) and distorted (solid, blue) power spectrum and correlation function, including the effects of both the anisotropic non-linear broadening and the non-linear correction DNL2. The gray and blue bands correspond to the same parameter variations as in Fig.~\ref{fig:distmodel}. We have adopted $\beta_{F}=1.4$ and $b_{F}(1+\beta_{F})=-0.351$ (corresponding to $b_{F}(1+\beta_{F})=-0.336$ at $z=2.25$ \cite{2011JCAP...09..001S}).}
   \label{fig:distpowercorrelation}
   \end{center}
\end{figure}

Figure~\ref{fig:distmodel} illustrates the distortion models for varying parameter values. DC1 shows a clear distinction in the effect of the parameters, where $k_{\rm C}$ controls the scale of the suppression and $p_{\rm C}$ modifies the shape. In the case of DC2, the difference between varying $k_{\rm C}$ and $p_{\rm C}$ is more subtle on the scale of the suppression, indicating a degeneracy between the parameters. Figure~\ref{fig:distpowercorrelation} shows the effects of the distortion model DC2 on $P_{F}$ and the corresponding correlation function at $z=2.3$. The suppression of the power spectrum on large scales distorts the broadband shape of the correlation function, but does not change the position of the BAO peak.

\section{BAO fitting method}
\label{sec:baofitting}

We implement the models for the non-linear correction and the broadband distortion into the publicly available fitting software baofit\footnote{See Appendix~\ref{sec:public} for information on where to download the fitting software.}. Our method for fitting the BAO in the \Lya auto-correlation is largely the same as in previous analyses using BOSS data \cite{2013A&A...552A..96B,2013JCAP...04..026S,2013JCAP...03..024K,2014JCAP...05..027F,2015A&A...574A..59D}. The correlation function is decoupled into a smooth and a peak component,
\begin{equation}
\xi_{F}(r_\parallel,r_\perp,\alpha_{\parallel},\alpha_{\perp}) = \xi_{\rm smooth} (r_\parallel, r_\perp) + a_{\rm peak} \cdot \xi_{\rm peak} (\alpha_{\parallel} r_\parallel , \alpha_{\perp} r_\perp )\ .
\end{equation}
Decoupling the peak is necessary in BAO fitting when the broadband shape is insufficiently understood. Although the $k$-space distortion model developed in this paper greatly improves the modeling of broadband distortions from continuum fitting for a decoupled fit, it is not quite refined enough for an unbiased measurement of the BAO scale for a fit where the peak is coupled to the broadband component. The parameter $a_{\rm peak}$ controls the amplitude of the BAO peak and is kept fixed to its nominal value $a_{\rm peak}=1$. The anisotropic shift of the observed peak position relative to the fiducial peak position is described by the radial and transverse scale factors
\begin{equation}
\alpha_{\parallel} = \frac { \left[D_{H}(z_{\rm eff})/r_d\right] }{\left[D_{H}(z_{\rm eff})/r_{d}\right]_{\rm fid}}
\quad , \quad
\alpha_{\perp} = \frac { \left[D_{A}(z_{\rm eff})/r_d\right] }{\left[D_{A}(z_{\rm eff})/r_{d}\right]_{\rm fid}} ~,
\label{eq:alpha}
\end{equation}
where $z_{\rm eff}$ is the effective redshift of the BAO measurement, $D_{H}(z)=c/H(z)$ is the Hubble distance, $D_{A}(z)$ is the comoving angular diameter distance, and $r_{d}$ is the sound horizon at the drag epoch \cite{2014MNRAS.439...83A} with a measured value 147.49~Mpc reported by Planck \cite{2014A&A...571A..16P}.

Decoupling the correlation function into a smooth and a peak component for the BAO fitting means that we must perform the corresponding split in the model of the matter power spectrum. Because the anisotropic non-linear broadening model is mainly valid on BAO scales, we choose to apply the broadening to the peak only, such that
\begin{equation}
P_{\rm NL}(k,\mu_{k},z) = \exp(-k^{2}\Sigma^{2}(\mu_{k})/2)P_{\rm peak}(k,z)+P_{\rm smooth}(k,z)\ .
\end{equation}
The linear peak component is defined by
\begin{equation}
P_{\rm peak}(k,z)=P_{\rm L}(k,z)-P_{\rm smooth}(k,z)\ ,
\end{equation}
where the smooth component $P_{\rm smooth}$ is the linear matter power spectrum from CAMB with the BAO feature erased following the ``sideband'' method of \cite{2013JCAP...03..024K}. Tabulated versions of $P_{\rm L}(k,z_{\rm ref})$ and $P_{\rm smooth}(k,z_{\rm ref})$ are inputs in the fits.

The key difference in our approach to BAO fitting compared to previous \Lya BAO analyses is the treatment of the broadband distortion. Using the $k$-space distortion model with two free parameters rather than an $r$-space broadband function with nine (or more) free parameters, we obtain a better goodness-of-fit for the DR11 auto-correlation while significantly reducing the number of fit parameters.

A more technical difference lies in the method for evaluating the correlation function. The first efforts for fitting BAO in the \Lya auto-correlation \cite{2013A&A...552A..96B,2013JCAP...04..026S,2013JCAP...03..024K} and the quasar-\Lya cross-correlation \cite{2014JCAP...05..027F} employed a method that relied on pre-calculated templates for the $l=0,2,4$ multipoles (monopole, quadrupole and hexadecapole) of the correlation function that had fixed anisotropic non-linear broadening applied to the BAO peak using an approximation valid for $\beta_{F}=1.4$. An improved method was established in \cite{2015A&A...574A..59D}, in which the BAO modeling was performed in $k$-space by means of the power spectrum of the transmitted flux fraction. This approach allows for exact application of the anisotropic non-linear broadening. For each set of model parameters, the correlation function was evaluated by first expanding $P_{F}$ into power spectrum multipoles $l=0,2,4$, and then performing a spherical Bessel transformation numerically of each multipole into its corresponding correlation function multipole. The non-linear broadening in principle transfers power to higher even multipoles ($l=6,8,...$), but the contribution from these higher-order multipoles to the correlation function is negligible. Instead of using a multipole expansion, we employ a direct 3D FFT of $P_{F}$ into the correlation function. This is because the distortion model transfers non-negligible power to all even multipoles and thus the sum of multipole contributions to the correlation function does not converge for the $l$ values ($l\lesssim14$) that can be robustly evaluated in the numerical spherical Bessel transformation.

The DR11 auto-correlation measurement uses 50 bins of width 4~Mpc/$h$, spanning the range from 0 to 200~Mpc/$h$, for each coordinate $r_{\parallel}$ and $r_{\perp}$. The total number of $(r_{\parallel},r_{\perp})$ bins is 2500. The fitting range is $40<r<180$~Mpc/$h$, reducing the number of data bins in the fits to 1515. We choose $b_{F}(1+\beta_{F})$ rather than $b_{F}$ to be one of the fit parameters, because it produces much less error correlation with $\beta_{F}$ \cite{2011JCAP...09..001S}. The six free parameters are: $\beta_{F}$, $b_{F}(1+\beta_{F})$, $\alpha_{\parallel}$, $\alpha_{\perp}$, $k_{\rm C}$ and $p_{\rm C}$.

\section{Tests on mock data sets}
\label{sec:mocks}

In this section, we test the BAO model by fitting it to the auto-correlation functions measured for DR11 mock data sets. Specifically, we compare the goodness-of-fit and the ability to recover the mock input values for the two distortion models. We start by introducing the mock samples that were generated for these tests.

\subsection{Mock data sets}
\label{subsec:mockdata}

The method for producing mock data sets with the BOSS DR11 survey geometry and simulated instrumental characteristics is described in \cite{2015JCAP...05..060B}. Following the method of \cite{2012JCAP...01..001F}, the absorption field for each mock sample was generated using an input $P_{F}$ with $\beta_{F}=1.4$ and $b_{F}(1+\beta_{F})=-0.336$ at $z=2.25$, and a cosmology matching our fiducial cosmology, with the exception of assuming massless neutrinos ($\Omega_{\rm \nu}h^{2}=0$). The redshift evolution of $P_{F}$ was assumed to follow equation~(\ref{eq:PFevo}). Non-linear effects were added to $P_{F}$ according to the model DNL1 with the parameter values specified in section~\ref{subsec:nlcorr}. Anisotropic non-linear broadening was not included for the mock samples.

We use two types of mock samples, referred to here as standard and noiseless mocks. The standard mocks were produced to emulate the characteristics of the real DR11 quasar spectra by adding instrumental noise and the effects of sky residuals, mis-calibration of the flux and mis-estimation of the noise from the BOSS data reduction pipeline to all the spectra \cite{2015JCAP...05..060B}. However, absorption from intervening damped \Lya (DLA) absorbers, Lyman limit systems (LLS) and metals were not included. A total of 100 independent realizations of the standard mock samples were produced. In addition, 100 noiseless mock samples using the same absorption field as the standard mocks but without any added instrumental noise or observational systematics were produced for the purpose of testing the BAO model and the fitting method under idealized conditions. 

Each mock sample was analyzed to measure the auto-correlation and covariance matrix as described in \cite{2015A&A...574A..59D}. The \Lya forest continua for the standard mocks were fitted using the ``C2" method. Since the standard mocks were previously used for testing BAO fitting for models with the $r$-space broadband function in \cite{2015A&A...574A..59D}, we can directly compare those results to the results obtained here using the $k$-space distortion model. For the noiseless mocks, we choose to use the true continua generated for each spectrum instead of performing the continuum fitting. The auto-correlations for the noiseless mocks thus do not feature broadband distortions ($D_{\rm C}=1$) and are used for cross-checking the consistency with the input values in the absence of distortions.

\subsection{Effective redshift correction}
\label{subsec:zeffcorr}

\begin{figure}
   \begin{center}
   \includegraphics[width=3.0in]{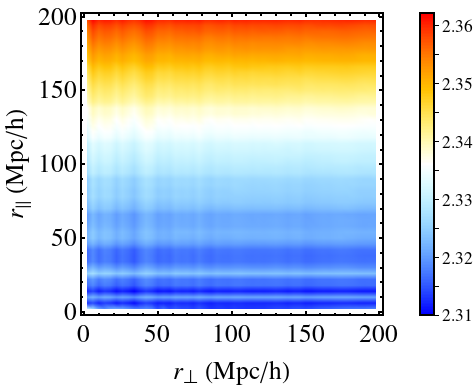}
   \includegraphics[width=3.0in]{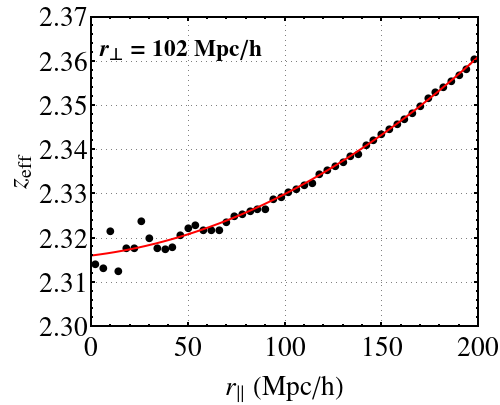}
   \caption{Variation of the effective redshift for the standard mocks. Left: Effective redshift for each bin $(r_{\parallel},r_{\perp})$. Right: Effective redshift as a function of $r_{\parallel}$ for $r_{\perp}=102$~Mpc/$h$. The red solid line shows the best fit using equation~(\ref{eq:zeff}).}
   \label{fig:zeff}
   \end{center}
\end{figure}

We adhere to the definition of the (overall) effective redshift of the BAO measurement as the average redshift of all the pixel pairs contributing to the correlation function in the region $80<r<120$~Mpc/$h$. The values are $z_{\rm eff}=2.322$ for the standard mocks and $z_{\rm eff}=2.244$ for the noiseless mocks. The difference is due to the different noise and added observational systematics that change the weighting of the pixel pairs. Figure~\ref{fig:zeff} shows the effective redshift for each bin in the $(r_{\parallel},r_{\perp})$ plane for the standard mocks. The value of $z_{\rm eff}$ increases in the $r_{\parallel}$ direction, but shows negligible change in the $r_{\perp}$ direction. This variation along the line of sight is due to the increasing relative contribution from quasars at higher redshift which have longer forest lengths. If this effect is not corrected for in the fits, it leads to percent-level systematic offsets in the values of $\beta_{F}$ and $b_{F}(1+\beta_{F})$.

We model the variation of the effective redshift using a second-order polynomial in $r_{\parallel}$,
\begin{equation}
z_{\rm eff}(r_{\parallel})=z_{0}+z_{1}\left( \frac{r_{\parallel}}{100\ {\rm Mpc}/h} \right)+z_{2}\left( \frac{r_{\parallel}}{100\ {\rm Mpc}/h} \right)^{2}\ .
\label{eq:zeff}
\end{equation}
The right panel of figure~\ref{fig:zeff} gives the effective redshift for the slice along $r_{\perp}=102$~Mpc/$h$. Equation~(\ref{eq:zeff}) provides a good fit to the data points. We repeat the same exercise for the noiseless mocks as well as for the DR11 data. The best-fit coefficients $z_{0}$, $z_{1}$ and $z_{2}$ are summarized in table~\ref{table:zeff}.

\begin{table}[htb]
\begin{center}
\begin{tabular}{|l|ccc|}
\hline
Sample & $z_{0}$ & $z_{1}$ & $z_{2}$\\
\hline
noiseless mocks & 2.1832 & 0.0221 & 0.0068 \\
standard mocks & 2.3161 & 0.0053 & 0.0086 \\
DR11 data & 2.3245 & 0.0051 & 0.0080 \\
\hline
\end{tabular}
\end{center}
\caption{Best-fit coefficients for the effective redshift correction of equation~(\ref{eq:zeff}) for the noiseless and standard mock samples and the DR11 data.}
\label{table:zeff}
\end{table}

\subsection{BAO fitting of mock data sets}
\label{subsec:mockfits}

We perform fits to the auto-correlation functions for 100 noiseless mock samples without distortion modeling ($D_{\rm C}=1$) and for 100 standard mock samples using distortion models DC1 and DC2. All the fits feature the DNL1 model for non-linear corrections, no anisotopic non-linear broadening ($\Sigma_{\parallel}=\Sigma_{\perp}=0$) and a linear matter power spectrum calculated for the mock cosmology at $z_{\rm ref}=2.25$, in accordance with the model for $P_{F}$ of the mocks. The goal of this exercise is to recover the fiducial values $\alpha_{\parallel}=1$, $\alpha_{\perp}=1$, $\beta_{F}=1.4$ and $b_{F}(1+\beta_{F})=-0.336$ at $z=2.25$, on average.

Starting with the results for the 100 noiseless mocks, table~\ref{table:mocksfits} summarizes the unweighted mean values with 1$\sigma$ errors for each parameter. While we are able to recover the correct mean value of $\alpha_{\parallel}$, we notice a small but statistically significant positive offset in the mean value of $\alpha_{\perp}$. This offset represents a systematic error that is not due to broadband modeling, spectral noise or the decoupling of the peak, and persists when using the fitting method of \cite{2015A&A...574A..59D}\footnote{This systematic error was not discovered in \cite{2015A&A...574A..59D} where only standard mocks were analyzed.}. There are a couple of approximations in the generation of the mocks that could explain the offset. First, the code to generate the mocks is based on the assumption that the different lines of sight are parallel. In order to add the correct geometry in our mock skewers, we have to generate all the skewers at different redshifts (four in the mocks used for this analysis), and use a linear interpolation as a function of redshift to generate a value for each pixel. Second, the code to generate the mocks uses the small angle approximation to compute the transverse separations between two pixels. The analysis code, however, uses a better approximation that is correct at second order in the angular separation. A sub-percent systematic error for $\alpha_{\perp}$ is less than one fifth of the statistical uncertainty reported in our DR11 analysis, and therefore can be regarded as a satisfactory performance. The recovered mean values of $\beta_{F}$ and $b_{F}(1+\beta_{F})$ are consistent with their fiducial values. We note that fits with the distortion model included yield $D_{\rm C}\rightarrow1$ ($k_{\rm C}\rightarrow0$), as expected, but its built-in property of vanishing at $k_{\parallel}=0$ introduces a bias in the recovered parameters.

\begin{table}
\begin{center}
\begin{tabular}{|l|cccc|}
\hline
Analysis & $\alpha_{\parallel}$ & $\alpha_{\perp}$ & $\beta_{F}$ & $b_{F}(1+\beta_{F})$\\
\hline
fiducial & $1.0$ & $1.0$ & $1.4$ & $-0.336$ \\
noiseless mocks & $0.9993 \pm 0.0014$ & $1.0064 \pm 0.0023$ & $1.390 \pm 0.007$ & $-0.3365 \pm 0.0003$ \\
std. mocks, DC1 & $1.0067 \pm 0.0029$ & $0.9984 \pm 0.0054$ & $1.493 \pm 0.013$ & $-0.3216 \pm 0.0005$ \\
std. mocks, DC2 & $1.0047 \pm 0.0029$ & $1.0030 \pm 0.0054$ & $1.397 \pm 0.009$ & $-0.3347 \pm 0.0004$ \\
\hline
\end{tabular}
\end{center}
\caption{Summary of the results for fits to 100 noiseless mock samples and 100 standard mock samples. The parameter values shown are unweighted mean values with $1\sigma$ errors and should be compared to the fiducial values that were used for generating the mocks. DC2 uses the fixed value $p_{\rm C}=3$, as motivated in the text.}
\label{table:mocksfits}
\end{table}

\begin{figure}
   \begin{center}
   \includegraphics[width=2.8in]{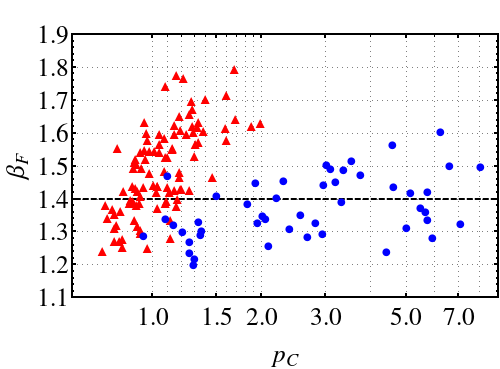}
   \includegraphics[width=3.0in]{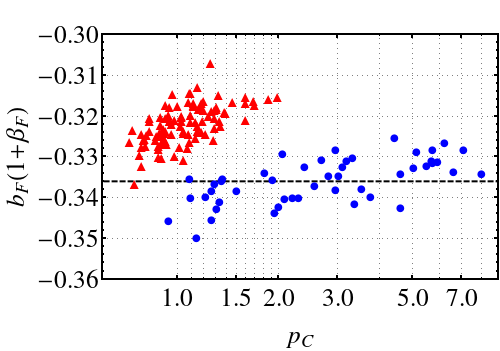}
   \includegraphics[width=3.0in]{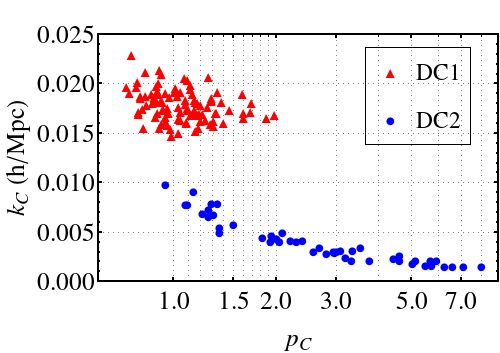}
   \caption{Best-fit parameter values for $\beta_{F}$, $b_{F}(1+\beta_{F})$ and $k_{\rm C}$ as a function of $p_{\rm C}$ for fits to the standard mocks. Results are shown for 100 mock samples for DC1 (red triangles) and 48 mock samples for DC2 (blue dots). The black dashed lines indicate the fiducial values that were used for generating the mocks. Errors bars have been omitted for clarity.}
   \label{fig:baoparampc}
   \end{center}
\end{figure}

For the analysis of the 100 standard mocks, we begin by investigating the performances of the distortion models with respect to $\beta_{F}$ and $b_{F}(1+\beta_{F})$. Figure~\ref{fig:baoparampc} displays the best-fit parameter values of $\beta_{F}$, $b_{F}(1+\beta_{F})$ and $k_{\rm C}$ as a function of $p_{\rm C}$. Both distortion models exhibit a correlation where $\beta_{F}$ and $b_{F}(1+\beta_{F})$ increase with $p_{\rm C}$. For DC2, the degeneracy between $k_{\rm C}$ and $p_{\rm C}$ causes the fits to 52 of the mock samples to fail to converge. The degeneracy is evident in the bottom panel where there is a clear correlation between the distortion parameters. Moving forward, we fix the $p_{\rm C}$ parameter for DC2 in order to remove this degeneracy to improve stability and performance of the fits. The value $p_{\rm C}=3$ makes both $\beta_{F}$ and $b_{F}(1+\beta_{F})$ agree with their fiducial values. DC1 does not suffer from the same degeneracy, and there is no value of $p_{\rm C}$ that produces agreement for both $\beta_{F}$ and $b_{F}(1+\beta_{F})$, so we will continue to treat it as a two-parameter model.

The main results for the 100 standard mocks are presented in table~\ref{table:mocksfits}. Both distortion models give mean values of $\alpha_{\parallel}$ and $\alpha_{\perp}$ that are statistically consistent with the fiducial values (although also consistent with the offset value found for the noiseless mocks). It is reassuring to find that neither distortion model biases the BAO peak position. The corresponding values obtained for BAO models with the $r$-space broadband function in \cite{2015A&A...574A..59D} are $\alpha_{\parallel}=1.003\pm0.003$ and $\alpha_{\perp}=1.008\pm0.006$, in agreement with our results. DC2 is also successful in recovering values of $\beta_{F}$ and $b_{F}(1+\beta_{F})$ that agree with the fiducial ones to 0.2\% and 0.4\%, respectively. We emphasize that similar results are obtained for DC2 even if $p_{\rm C}$ is a free parameter (see figure~\ref{fig:baoparampc}). DC1, on the other hand, yields systematically higher values by about 7\% for $\beta_{F}$ and 4\% for $b_{F}(1+\beta_{F})$. We have additionally performed fits to the correlation function obtained from stacking all the standard mocks and find parameter results consistent with those in table~\ref{table:mocksfits}.

\begin{figure}[t]
   \begin{center}
   \includegraphics[width=2.5in]{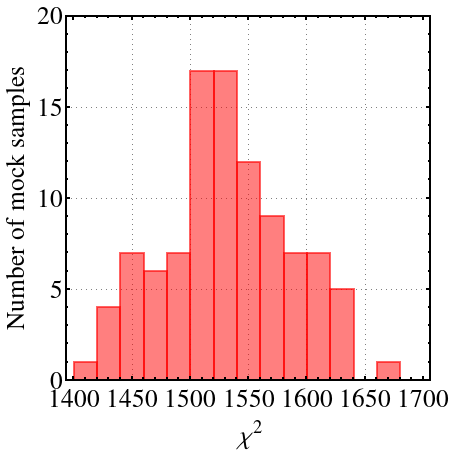}
   \includegraphics[width=2.5in]{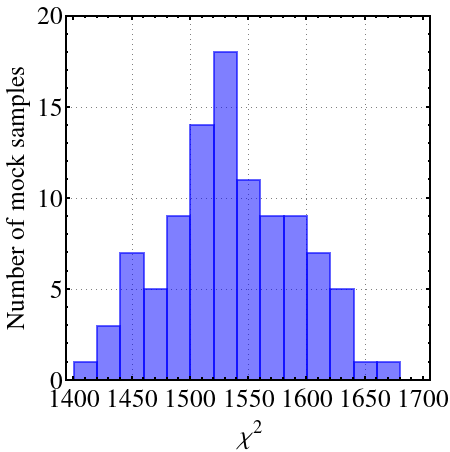}
   \caption{Distribution of the minimum $\chi^2$ values for fits to the 100 standard mocks using DC1 (left) and DC2 with $p_{\rm C}=3$ (right). Mean values are $\chi^{2}=1531.6$ for 1509 degrees of freedom for DC1, and $\chi^{2}=1534.7$ for 1510 degrees of freedom for DC2.}
   \label{fig:chi2}
   \end{center}
\end{figure}

Figure~\ref{fig:chi2} presents histograms of the minimum $\chi^{2}$ values for fits to the standard mocks. The mean values are comparable, with $\chi^{2}=1531.6$ for 1509 degrees of freedom for DC1, and $\chi^{2}=1534.7$ for 1510 degrees of freedom for DC2. This is a sound result in support of the distortion modeling, and a significant improvement over the mean value $\chi^{2}=1572.8$ for 1502 degrees of freedom obtained for the BAO model with the $r$-space broadband function in \cite{2015A&A...574A..59D}.

\begin{figure}[t]
   \begin{center}
   \includegraphics[width=6.0in]{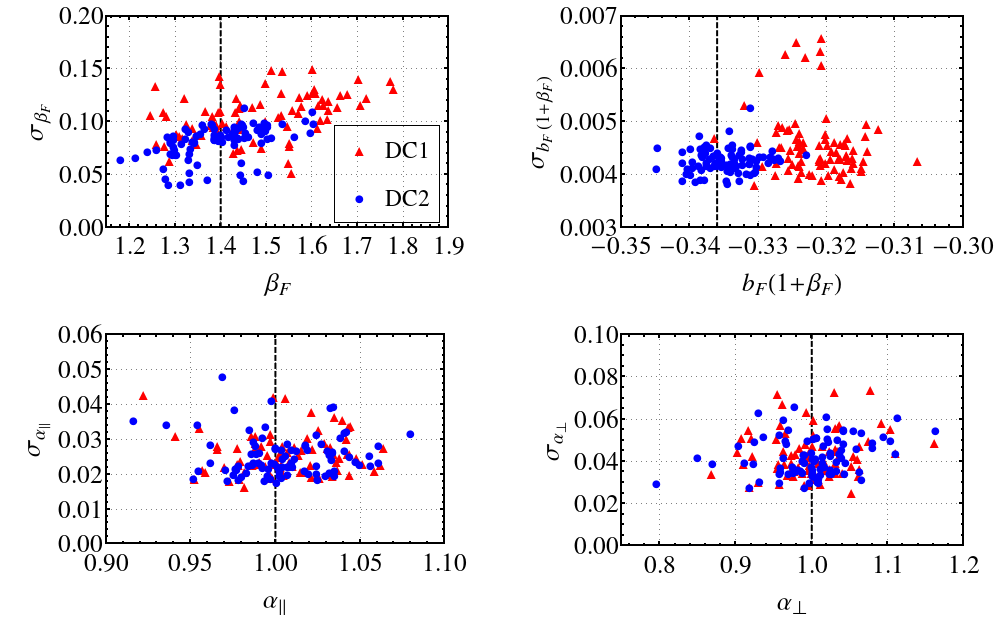}
   \caption{Best-fit parameter values and errors for fits to the standard mocks using DC1 (red triangles) and DC2 with $p_{\rm C}=3$ (blue dots). Results are shown for 84 and 93 mock samples with accurate parameter covariance matrices for DC1 and DC2, respectively. The black dashed lines indicate the fiducial values that were used for generating the mocks.}
   \label{fig:baoparamerr}
   \end{center}
\end{figure}

The best-fit parameter values and their 1$\sigma$ errors are displayed in figure~\ref{fig:baoparamerr}. Results are presented for 84 mock samples for DC1 and 93 mock samples for DC2, for which the fit was able to accurately estimate the parameter covariance matrices. The data points are scattered symmetrically around the fiducial values for DC2, whereas DC1 produces an offset for $\beta_{F}$ and $b_{F}(1+\beta_{F})$. The parameter errors are quite comparable between the models, and their mean values for DC2 are $\sigma_{\beta_{F}}=0.081$, $\sigma_{b_{F}(1+\beta_{F})}=0.0043$, $\sigma_{\alpha_{\parallel}}=0.025$ and $\sigma_{\alpha_{\perp}}=0.042$.

\begin{figure}[htb]
   \begin{center}
   \includegraphics[width=3.0in]{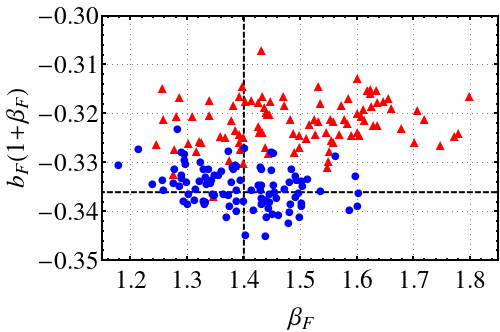}
   \includegraphics[width=3.0in]{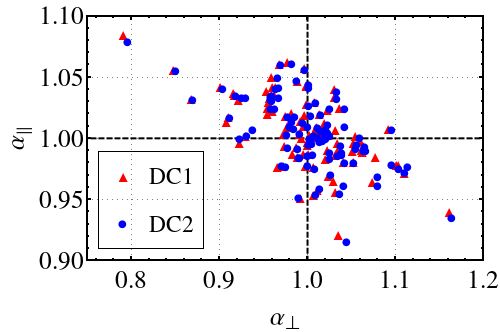}
   \caption{Plots of the best-fit parameter values for fits to the 100 standard mocks using DC1 (red triangles) and DC2 with $p_{\rm C}=3$ (blue dots). The black dashed lines indicate the fiducial values that were used for generating the mocks. Error bars have been omitted for clarity.}
   \label{fig:baoparamscatter}
   \end{center}
\end{figure}

Finally, figure~\ref{fig:baoparamscatter} shows scatter plots for the best-fit parameters. The data points are aligned horizontally for $\beta_{F}$ and $b_{F}(1+\beta_{F})$, indicating that these parameters are uncorrelated. The anticorrelation between the recovered $\alpha_{\parallel}$ and $\alpha_{\perp}$ agrees with the result of \cite{2015A&A...574A..59D}. Comparing the values on a mock-by-mock basis, $\alpha_{\parallel}$ is typically higher for DC1 by 0.1\% and $\alpha_{\perp}$ is typically higher for DC2 by 0.3\%, but larger differences on the percent-level occur for a few mocks.

\section{Results}
\label{sec:results}

Fits to the DR11 auto-correlation feature the DNL2 model for non-linear corrections, fixed anisotropic non-linear broadening of the BAO peak and a linear matter power spectrum calculated for the fiducial cosmology at $z_{\rm ref}=2.3$. The effective redshift of the measurement is $z_{\rm eff}=2.34$ and its correction in the $r_{\parallel}$ direction is described by equation~(\ref{eq:zeff}) using the coefficients listed in table~\ref{table:zeff}. Based on the results for the standard mocks in section~\ref{subsec:mockfits}, the baseline fit for deriving our main results uses the distortion model DC2 with $p_{\rm C}=3$.

\begin{figure}
   \begin{center}
   \includegraphics[width=3.0in]{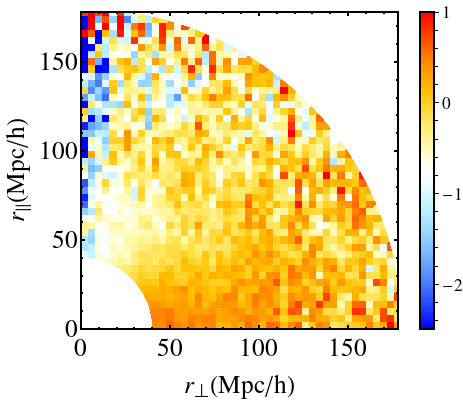}
   \includegraphics[width=3.0in]{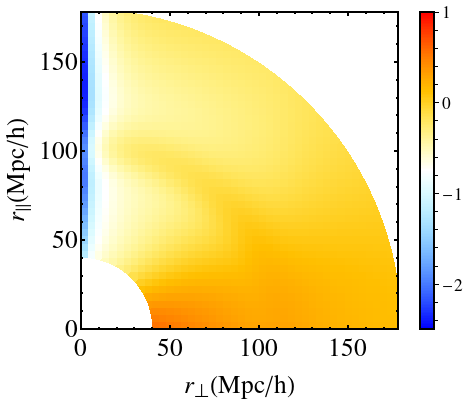}
   \caption{Measured DR11 auto-correlation $r^2\xi(r_{\parallel},r_{\perp})$ (left panel) and best-fit model using DC2 with $p_{\rm C}=3$ (right panel), as a function of radial ($r_{\parallel}$) and transverse ($r_{\perp}$) separations. The color scale saturates at $r^2\xi_{F}=1$ and $r^2\xi_{F}=-2.5$.}
   \label{fig:dr11data}
   \end{center}
\end{figure}

Figure~\ref{fig:dr11data} displays the DR11 auto-correlation measurement in the ($r_{\parallel}$,$r_{\perp}$) plane for the fitting range $40<r<180$~Mpc/$h$, along with the best-fit BAO model for the baseline fit. Table~\ref{table:bestfits} presents the best-fit results for the baseline fit along with several alternative fits. Parameter errors are estimated assuming the likelihood function is Gaussian at $1\sigma$ (a more rigorous error estimation for the baseline fit follows later in this section). The last row of the table (denoted ``Delubac15") is the results for the standard fit with the $r$-space broadband function in \cite{2015A&A...574A..59D} and serves as the main comparison. Both distortion models provide significantly better fits to the data, despite using fewer free parameters. The results for $\alpha_{\parallel}$ and $\alpha_{\perp}$ are consistent between all models, confirming that the measurement of the BAO peak position is robust to the different treatments of the broadband distortion. The largest benefit of using the distortion models is the much more precise constraints obtained for $\beta_{F}$ and $b_{F}(1+\beta_{F})$. The magnitudes of the statistical error estimated for the best fit for these parameters are reduced by a factor of five when using DC2. Compared to the best-fit values for DC2, DC1 yields values that deviate by about 6\% for $\beta_{F}$ and by 3\% for $b_{F}(1+\beta_{F})$, compatible with the magnitude of the systematic errors estimated for this model in section~\ref{subsec:mockfits}. Two of the alternative fits vary the assumptions for the baseline fit. Allowing $p_{\rm C}$ to be a free parameter does not improve the goodness-of-fit or change the best-fit parameters notably; the reason is that the best-fit value $p_{\rm C}=2.57\pm0.27$ is not far from $p_{\rm C}=3$. Ignoring the non-linear correction ($D_{\rm NL}=1$) produces no significant difference either.

\begin{table}
\begin{center}
\begin{tabular}{|l|ccccc|}
\hline
Analysis & $\alpha_{\parallel}$ & $\alpha_{\perp}$ & $\beta_{F}$ & $b_{F}(1+\beta_{F})$ & $\chi^2/DOF$\\
\hline
DC2 & $1.055 \pm 0.037$ & $0.963 \pm 0.062$ & $1.39 \pm 0.10$ & $-0.374 \pm 0.005$ & $1484.2/1510$ \\
\hline
free $p_{\rm C}$ & $1.054 \pm 0.037$ & $0.963 \pm 0.062$ & $1.37 \pm 0.09$ & $-0.375 \pm 0.006$ & $1484.2/1509$ \\
$D_{\rm NL}=1$ & $1.055 \pm 0.036$ & $0.963 \pm 0.061$ & $1.33 \pm 0.09$ & $-0.377 \pm 0.005$ & $1481.5/1510$ \\
\hline
DC1 & $1.057 \pm 0.032$ & $0.962 \pm 0.059$ & $1.31 \pm 0.09$ & $-0.363 \pm 0.007$ & $1484.7/1509$ \\
Delubac15 & $1.054 \pm 0.031$ & $0.973 \pm 0.051$ & $1.50 \pm 0.47$ & $-0.402 \pm 0.024$ & $1499.1/1502$ \\
\hline
\end{tabular}
\end{center}
\caption{Best-fit results for fits to the DR11 auto-correlation. Our baseline fit uses DC2 with $p_{\rm C}=3$ and is presented on the first row. Delubac15 refers to the standard fit in \cite{2015A&A...574A..59D} that uses an $r$-space broadband function to accommodate the broadband distortions.}
\label{table:bestfits}
\end{table}

\begin{figure}
   \begin{center}
   \includegraphics[width=3.5in]{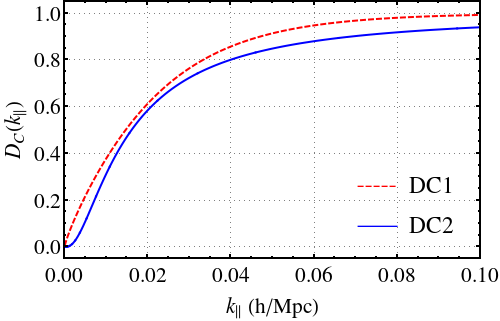}
   \caption{Best-fit distortion models DC1 (red dashed) and DC2 with $p_{\rm C}=3$ (blue solid) for fits to the DR11 auto-correlation. The best-fit distortion parameters are $k_{\rm C}=0.030\pm0.002$~$h$/Mpc and $p_{\rm C}=0.847\pm0.075$ for DC1, and $k_{\rm C}=0.0050\pm0.0001$~$h$/Mpc for DC2.}
   \label{fig:distmodels}
   \end{center}
\end{figure}

The best-fit distortion models are presented in figure~\ref{fig:distmodels}, with the parameter values specified in the figure caption. The values of $k_{\rm C}$ are larger than those found for the standard mocks, which suggests that there are additional sources of broadband distortion present in the data that are absorbed by the distortion models.

\begin{figure}
   \begin{center}
   \includegraphics[width=3.0in]{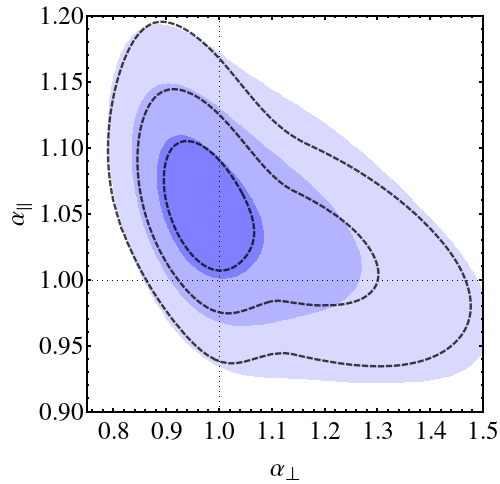}
   \caption{Constraints on $\alpha_{\parallel}$ and $\alpha_{\perp}$. Contours show 68.3\% ($\Delta\chi^{2}=2.30$), 95.45\% ($\Delta\chi^{2}=6.18$) and 99.73\% ($\Delta\chi^{2}=11.83$) confidence contours from the DR11 auto-correlation using the distortion model DC2 with $p_{\rm C}=3$. The overlayed dashed contours are for the standard fit in \cite{2015A&A...574A..59D} that uses an $r$-space broadband function to accommodate the broadband distortions. Results are in good agreement.}
   \label{fig:alphacontours}
   \end{center}
\end{figure}

\begin{figure}
   \begin{center}
   \includegraphics[width=4.0in]{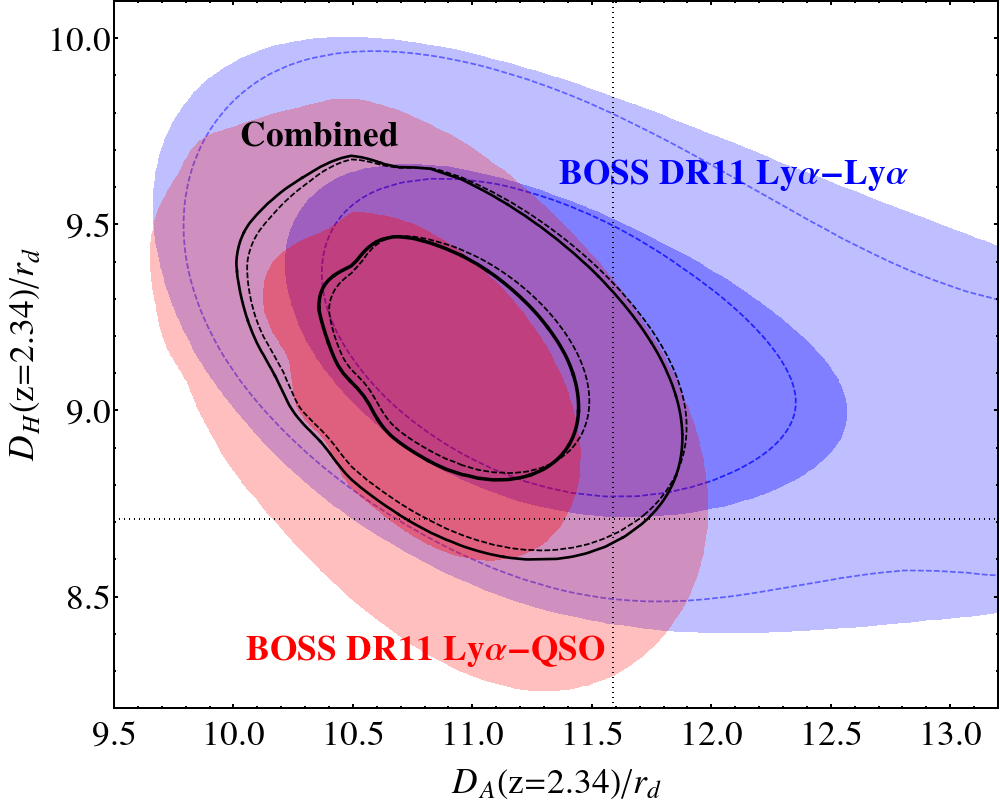}
   \caption{Constraints on $D_{H}/r_{d}$ and $D_{A}/r_{d}$. Contours show 68.3\% ($\Delta\chi^{2}=2.30$) and 95.45\% ($\Delta\chi^{2}=6.18$) confidence contours from the DR11 auto-correlation using the distortion model DC2 with $p_{\rm C}=3$ (blue), the DR11 quasar-\Lya cross-correlation \cite{2014JCAP...05..027F} (red), and the combined constraints (black). The dashed contours are the results of \cite{2015A&A...574A..59D}, in agreement with our constraints. The dotted lines indicate the fiducial values $D_{H}/r_{d}=8.708$ and $D_{A}/r_{d}=11.59$ at $z=2.34$.}
   \label{fig:combinedcontours}
   \end{center}
\end{figure}

Figure~\ref{fig:alphacontours} displays the 68.3\% ($\Delta\chi^{2}=2.30$), 95.45\% ($\Delta\chi^{2}=6.18$) and 99.73\% ($\Delta\chi^{2}=11.83$ confidence contours for the scale factors $\alpha_{\parallel}$ and $\alpha_{\perp}$ for the baseline fit, and compares them to the corresponding contours obtained in \cite{2015A&A...574A..59D}. The results are in good agreement, although the distortion model produces slightly larger contours that are more Gaussian in shape. Therefore, the constraints on the Hubble parameter and the angular diameter distance reported in \cite{2015A&A...574A..59D}, and consequently the cosmological implications, are unchanged. This conclusion is further illustrated in figure~\ref{fig:combinedcontours}, where we combine the auto-correlation contours with the contours from the quasar-\Lya cross-correlation \cite{2014JCAP...05..027F}. The constraints on $\alpha_{\parallel}$ and $\alpha_{\perp}$ have been converted into constraints on $D_{H}/r_{d}$ and $D_{A}/r_{d}$ using the fiducial values $D_{H}/r_{d}=8.708$ and $D_{A}/r_{d}=11.59$ at $z=2.34$. Our combined contours are in excellent agreement with the result of \cite{2015A&A...574A..59D}.

\begin{figure}
   \begin{center}
   \includegraphics[width=3.0in]{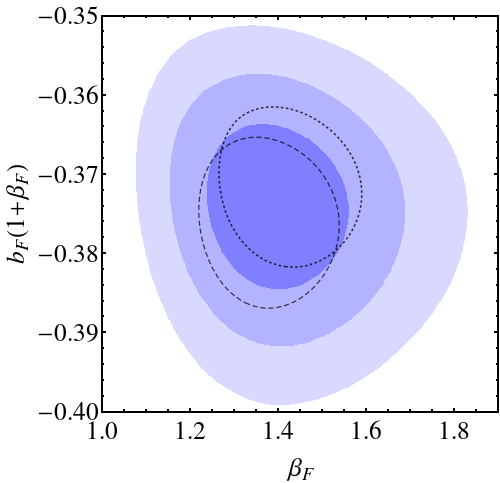}
   \caption{Constraints on $\beta_{F}$ and $b_{F}(1+\beta_{F})$ at $z=2.3$. Contours show 68.3\% ($\Delta\chi^{2}=2.30$), 95.45\% ($\Delta\chi^{2}=6.18$) and 99.73\% ($\Delta\chi^{2}=11.83$) confidence contours from the DR11 auto-correlation using the distortion model DC2 with $p_{\rm C}=3$. The overlayed 68.3\% confidence contours are for $p_{\rm C}=2.5$ (dashed) and $p_{\rm C}=4$ (dotted). Scale factors have been fixed to their best-fit values $\alpha_{\parallel}=1.055$ and $\alpha_{\perp}=0.963$.}
   \label{fig:betabiascontours}
   \end{center}
\end{figure}

Figure~\ref{fig:betabiascontours} shows the 68.3\%, 95.45\% and 99.73\% confidence contours for $\beta_{F}$ and $b_{F}(1+\beta_{F})$ at $z=2.3$ for the baseline fit. The marginalized 68.3\% ($\Delta\chi^{2}=1$), 95.45\% ($\Delta\chi^{2}=4$) and 99.73\% ($\Delta\chi^{2}=9$) constraints are
\begin{equation}
\beta_{F}=1.39^{+0.11\ +0.24\ +0.38}_{-0.10\ -0.19\ -0.28}
\quad , \quad
b_{F}(1+\beta_{F})=-0.374^{+0.007\ +0.013\ +0.020}_{-0.007\ -0.014\ -0.022}\ .
\end{equation}
In comparison, the marginalized 68.3\% constraints for the Delubac15 fit are $\beta_{F}=1.47^{+1.70}_{-0.72}$ and $b_{F}(1+\beta_{F})=-0.400^{+0.055}_{-0.050}$, and thus the statistical errors are reduced by more than a factor of seven for our baseline fit. The overlayed 68.3\% confidence contours in figure~\ref{fig:betabiascontours} demonstrate how the constraints vary when changing to $p_{\rm C}=4$ and $p_{\rm C}=2.5$. These values yielded acceptable agreement for $\beta_{F}$ and $b_{F}(1+\beta_{F})$, respectively, for the standard mocks. This illustrates the (maximum) level of systematic error associated with the distortion model. Increasing $p_{\rm C}$ moves $\beta_{F}$ and $b_{F}(1+\beta_{F})$ to higher values, whereas decreasing $p_{\rm C}$ gives lower values, echoing the correlation observed in the results for the standard mocks (see figure~\ref{fig:baoparampc}). The value $\beta_{F}=1.4$ assumed for the mock data sets and predicted in hydrodynamical simulations \cite{2015arXiv150604519A} agrees well with our measurement. Interestingly, $\beta_{F}=0.71^{+0.21}_{-0.16}$ and $b_{F}(1+\beta_{F})=-0.351\pm0.013$ (corresponding to $b_{F}(1+\beta_{F})=-0.336$ at $z=2.25$; assumed for the mock data sets) reported by \cite{2011JCAP...09..001S} for smaller scales $20<r<100$~Mpc/$h$ are in moderate tension with our constraints. We also note that the best-fit value $\beta_{F}=1.12^{+0.32}_{-0.26}$ for the quasar-\Lya cross-correlation \cite{2014JCAP...05..027F} is compatible within 68\% confidence with our measurement.

Lastly, the measured values of $\beta_{F}$ and $b_{F}(1+\beta_{F})$ at $z=2.3$, with a correlation coefficient of -0.183, can be converted through error propagation into constraints on the \Lya forest bias parameter
\begin{equation}
b_{F} = -0.157\pm 0.007\ ,
\end{equation}
and the \Lya forest velocity bias parameter $b_{v}=b_{F}\beta_{F}/f$, 
\begin{equation}
b_{v} = -0.226\pm 0.009\ ,
\end{equation}
where we have adopted $f=0.961$ at this redshift for our fiducial cosmology. These measured values of $b_{F}$ and $b_{v}$ are somewhat lower (higher in absolute value) than predictions from hydrodynamical simulations \cite{2003ApJ...585...34M,2015arXiv150604519A} and theoretical models \cite{2012JCAP...03..004S}.

\section{Discussion}
\label{sec:discussion}

In this paper, we have described a $k$-space model for the broadband distortion of the three-dimensional \Lya forest auto-correlation function induced by quasar continuum fitting. The distortion model improves on the previous treatment of broadband distortions in BAO fitting by providing a lower $\chi^{2}$ using fewer parameters for the fit to the BOSS DR11 \Lya auto-correlation. More importantly, it enables a measurement of the \Lya forest redshift-space distortion parameter $\beta_{F}$ to a statistical precision of 8\% and the combination with the linear bias parameter $b_{F}(1+\beta_{F})$ to 2\%, a reduction of the statistical errors by more than a factor of seven. The systematic errors on these parameters associated with the distortion model are estimated using mock data sets to be less than 0.5\%. Regarding the BAO result, we confirm that the measurement of the peak position is robust to the different treatments of the broadband distortion. Our results for $\alpha_{\parallel}$ and $\alpha_{\perp}$ are in good agreement with \cite{2015A&A...574A..59D}, and thus the cosmological implications of the DR11 \Lya BAO measurement remain the same.

Even though broadband distortions are primarily due to continuum fitting, additional distortions of the auto-correlation could arise from instrumental systematics (spurious correlations from, e.g., sky residuals and spectro-photometric calibration introduced by the data reduction pipeline), large-scale intensity fluctuations in the cosmic ionizing background radiation \cite{2014PhRvD..89h3010P,2014arXiv1404.7425G}, large-scale temperature fluctuations in the intergalactic medium \cite{2015MNRAS.447.2503G}, and absorption by high column density systems and metals \cite{2015JCAP...05..060B,2012JCAP...07..028F}. The fact that we obtain a good fit to the auto-correlation with the current BAO model suggests that these effects are either small, serendipitously cancel each other, or are absorbed by the distortion model. Regardless of the reason, it is important for future studies to better understand how these effects modify the auto-correlation and then correct for any significant contaminations and refine the BAO model to include the relevant astrophysical effects. Improvements of the data reduction pipeline are anticipated for upcoming BAO analyses \cite{2015arXiv150604790M} and should alleviate or remove potential instrumental systematics in the measurement of the BAO peak position and the broadband shape. While DLAs are already accounted for in the analysis, absorber systems with lower column densities are difficult to reliably identify in noisy spectra and thus remain a potential source of systematic error. The effect of such absorbers on the auto-correlation is to decrease the measured values of $\beta_{F}$ and $b_{F}$ by a few percent \cite{2012JCAP...07..028F} and was proposed in \cite{2011JCAP...09..001S} as an explanation for the discrepancy between the measured $\beta_{F}$ and theoretical predictions from numerical simulations of the \Lya forest \cite{2003ApJ...585...34M,2015arXiv150604519A}. The scale dependence introduced by fluctuations in the ionizing background leads to a suppression of the power spectrum on large scales. Since this result is degenerate with the continuum fitting distortion, it can be absorbed by the distortion model \cite{2014ApJ...792L..34P}.

Given the improved modeling of the broadband shape, it is desirable to extend the fitting range to smaller separations ($r<40$~Mpc/$h$) and thereby further reduce the statistical errors on $\beta_{F}$ and $b_{F}(1+\beta_{F})$. Such a fit finds lower $\beta_{F}$ and higher $b_{F}(1+\beta_{F})$, changes in the directions of the values reported in \cite{2011JCAP...09..001S}. However, the distortion models considered in our analysis do not give unbiased measurements for the larger fitting range. The systematic errors on $\beta_{F}$ and $b_{F}(1+\beta_{F})$ increase to 2-3\% for fits with $r>20$~Mpc/$h$, roughly the same magnitude as the statistical errors, and thus a more careful study involving modifications to the distortion model parameterizations is needed to improve on this result. Furthermore, the use of smaller scales would require more attention to the non-linear modeling. The 3D FFT resolution employed in this analysis restricts the wavenumber range to $k<0.79$~$h$/Mpc, which omits the contribution from the small-scale power spectrum in the calculation of the correlation function. Another interesting possibility that requires further refining of the broadband modeling is to constrain the scale factors $\alpha_{\parallel}$ and $\alpha_{\perp}$ using the full shape of the correlation function in the Alcock-Paczy\'nski test \cite{1979Natur.281..358A}. Fits without decoupling of the peak from the broadband component could potentially reduce the statistical errors on the measured scale factors by a factor of two.

We expect the general approach of the distortion model to also apply for the quasar-\Lya cross-correlation, albeit with some modifications. Since the length of the \Lya forest is a function of redshift, and the contributions to the cross-correlation function for positive radial separations (absorption behind the quasar) are produced by spectra with predominantly higher redshifts compared to negative radial separations, the broadband distortion becomes asymmetric with respect to $r_{\parallel}=0$ (see figure~2 of \cite{2014JCAP...05..027F}). We therefore anticipate that the distortion model needs to be generalized to accommodate this asymmetry, in order to achieve optimal results for BAO fitting of the cross-correlation.

The BOSS data set has opened the possibility to measure the expansion rate and geometry of the Universe at redshift $z>2$ using BAO in the \Lya forest. The precision of the BAO measurement, the level of sophistication in the fitting method and our understanding of the systematic errors have improved for each analysis of the growing data set, and the final BOSS data release \cite{2015ApJS..219...12A} (DR12) will provide the best measurement yet of the BAO scale at $z\simeq2.3$. The advances in broadband distortion modeling presented in this paper now also enable significantly more accurate measurements of the large-scale linear bias and redshift-space distortion of the \Lya forest.

\section*{Acknowledgments}

Funding for SDSS-III has been provided by the Alfred P. Sloan Foundation, the Participating Institutions, the National Science Foundation, and the U.S. Department of Energy Office of Science. The SDSS-III web site is \url{http://www.sdss3.org/}.

SDSS-III is managed by the Astrophysical Research Consortium for the Participating Institutions of the SDSS-III Collaboration including the University of Arizona, the Brazilian Participation Group, Brookhaven National Laboratory, Carnegie Mellon University, University of Florida, the French Participation Group, the German Participation Group, Harvard University, the Instituto de Astrofisica de Canarias, the Michigan State/Notre Dame/JINA Participation Group, Johns Hopkins University, Lawrence Berkeley National Laboratory, Max Planck Institute for Astrophysics, Max Planck Institute for Extraterrestrial Physics, New Mexico State University, New York University, Ohio State University, Pennsylvania State University, University of Portsmouth, Princeton University, the Spanish Participation Group, University of Tokyo, University of Utah, Vanderbilt University, University of Virginia, University of Washington, and Yale University.


\appendix
\section{Public access to data and code}
\label{sec:public}

The software used in this paper is available at \url{https://github.com/deepzot/baofit/}. We also provide instructions for how to install and run the software, together with the correlation-function estimates and configuration files necessary to reproduce our main results, at \\ \url{http://darkmatter.ps.uci.edu/baofit/}. The software is written in C++ and uses MINUIT~\cite{1975JamesRoos} for likelihood minimization. Data and configuration files are in plain text format.

\bibliographystyle{JHEP}
\bibliography{distortionmodelbib}

\end{document}